\documentclass{elsart}

\usepackage{natbib}

\usepackage{graphicx}
\usepackage{amssymb}

\def\parn{\par\noindent}
\def\ltsim{\raise 2pt \hbox {$<$} \kern-1.1em \lower 4pt \hbox {$\sim$}}
\def\ltapprox{\raise 2pt \hbox {$<$} \kern-1.1em \lower 5pt \hbox {$\approx
$}}
\def\gtsim{\raise 2pt \hbox {$>$} \kern-1.1em \lower 4pt \hbox {$\sim$}}
\def\gtapprox{\raise 2pt \hbox {$>$} \kern-1.1em \lower 5pt \hbox {$\approx
$}}
\def\astrobj#1{#1}

\begin{document}

\begin{frontmatter}



\title{The cluster relic source in \astrobj{A 521}}

\author[1,2,3]{S.~Giacintucci}
\author[2]{T.~Venturi}
\author[1]{S.~Bardelli}
\author[2]{G.~Brunetti}
\author[3]{R.~Cassano}
\author[2,3]{D.~Dallacasa}

\address[1]{INAF -- Osservatorio Astronomico di Bologna, 
via Ranzani 1, I--40126 Bologna, Italy}
\address[2]{INAF -- Istituto di Radioastronomia, via Gobetti 101, I-40129, Bologna, Italy}
\address[3]{Dipartimento di Astronomia, Universit\`a di Bologna, via Ranzani 1, I--40126, 
Bologna, Italy}

\begin{abstract}
We present high sensitivity radio observations of the 
merging cluster \astrobj{A 521}, at a mean redsfhit z=0.247. The observations 
were carried out with the GMRT at 610 MHz and cover a region of $\sim$1 square
degree, with a sensitivity limit of $1\sigma$ = 35 $\mu$Jy b$^{-1}$.
\\
The most relevant result of these observations is the presence of a radio 
relic 
at the cluster periphery, at the edge of a region where group infalling into
the main cluster is taking place.
Thanks to the wealth of information available in the literature 
in the optical and X--ray bands, a multi--band study of the relic and its 
surroundings was performed. Our analysis is suggestive of a connection 
between this source and the complex ongoing merger in the \astrobj{A 521} region.
The relic might be ``revived' fossil radio plasma through adiabatic compression
of the magnetic field or shock re--acceleration due to the merger events. 
We also briefly discussed the possibility that this source is the result of 
induced ram pressure stripping of radio lobes associated with the nearby cluster 
radio galaxy J0454--1016a.
\\
Allowing for the large uncertainties due to the small statistics, 
the number of radio emitting early--type galaxies found in \astrobj{A 521}  
is consistent with the expectations from the standard radio luminosity 
function for local 
(z$\le$0.09) cluster ellipticals.
\end{abstract}


\end{frontmatter}

% main text
\section{Introduction}
%\label{}

Radio observations reveal that a number of galaxy clusters 
host diffuse synchrotron radio emission, not obviously
associated with cluster galaxies, extended on cluster 
scale and referred to as {\it radio halos} and {\it relics}. 
These sources probe the presence of magnetic 
fields and relativistic particles mixed with the hot gas in 
the intracluster medium (ICM).
A promising approach  in our understanding of the nature 
of these sources is the
possibility that turbulence and shocks
induced by cluster mergers may be able to re--accelerate 
pre--existing electrons in the ICM, producing the emission 
from radio halos and relics (see the recent reviews by Brunetti
2003 and 2004, Sarazin 2004).

Both halos and relics are characterised by very low
surface brightness. They lack an obvious optical counterpart
and can reach sizes $\gtsim$ Mpc.
The class of radio halos is at present well defined (see 
Giovannini \& Feretti 2002 for a review).
Halos are detected in the central regions of galaxy clusters, 
show a fairly regular shape, and are usually unpolarized. 
They are characterised by steep integrated radio spectra, 
i.e. $\alpha$ \gtsim 1 for S $\propto \nu^{-\alpha}$, although
the spectral index distribution may 
show small scale inhomogeneities (Feretti et al. 2004).
High frequency spectral steepening is present in few cases 
(Feretti 2005, Giacintucci et al. 2005).
\\
Cluster relic sources are less homogeneous and more difficult to
classify, possibly due to our still limited knowledge 
and understanding of their formation and evolution. 
They are usually located in peripheral cluster regions, 
and show many different morphologies (sheet, arc, irregular, toroidal).
At present $\sim$ 20 relics and candidates are known
(Kempner \& Sarazin 2001; Giovannini \& Feretti 2004), 
however the observational information is still limited.
Their radio emission is usually highly polarized (up to 30\%).
For those few sources with multifrequency imaging, a steep
integrated spectrum is found ($\alpha$ \gtsim 1, up to ultra--steep
regimes). 
\\
\\ 
All clusters known to host a radio halo 
and/or a relic soure are found to show some degree of disturbance 
in the distribution of the hot gas and of the optical galaxies
(Buote 2001, Schuecker et al. 2001, 
Sarazin 2002). Some well studied and impressive examples are 
for instance the radio halo in \astrobj{A 2163} (Feretti et al. 2001) and
the double relics in \astrobj{A 3667} (Roettiger et al. 1999) and 
\astrobj{A 3367} (Bagchi et al. 2005).

It is interesting to point out that the observational link between 
cluster halos, relics and the merging phenomena has been outlined
{\it a posteriori}.
A different and promising approach is the {\it a priori} selection
of clusters experiencing well studied merging events, in order 
to determine the occurence of halos and relics.
As an example, deep radio observations of the merging
chain of clusters in the core of the Shapley Concentration 
led to the discovery of a radio halo at the centre of \astrobj{A 3562},
which is the faintest radio halo known to date
(Venturi et al. 2000 and 2003, Giacintucci et al. 2004 
and 2005), and
given its very low surface brightness it would have not been 
detected in a ``blind''  radio survey.

In this paper we present Giant Metrewave Radio Telescope
(GMRT) 610 MHz observations of
the galaxy cluster \astrobj{A 521}. This cluster was selected for the
search of extended cluster scale radio emission on the
basis of the wealth of optical (Maurogordato et al. 2000,
hereinafter M00; Ferrari et al. 2003, hereinafter F03)
and X--ray (Arnaud et al. 2000, hereinafter A00; 
Ferrari et al. 2005, hereinafter F05) information
available in the literature, suggesting a very complex dynamical state.
Furthermore, thanks to the amount of photometric and spectroscopic
data available,  \astrobj{A 521} is also
an ideal environment to study the effects of cluster mergers on the
radio emission properties of the member galaxy population, which is
still controversial (see e.g. Venturi et al. 2000,
Giacintucci et al. 2004, Owen et al. 1999,
Miller \& Owen 2003, Miller et al. 2003).

The observations presented here are part
of a much larger project carried out with a deep GMRT radio survey
at 610 MHz (Venturi et al. in prep.), whose aim is the search for radio
halos and relics in clusters at intermediate redshift (z=0.2$\div$0.4),
to test current statistical expectations from models for the formation 
of cluster radio halos (Cassano et al. 2004; Cassano \& Brunetti
2005; Cassano et al. in prep.).
\\
\\
The outline of the paper is as follows: in Section \ref{sec:a521} we 
give a summary of the main properties of \astrobj{A 521} and its complex dynamical 
state of merging; the 610 MHz observations and the data reduction are 
described in Section \ref{sec:obs}; in Section \ref{sec:sample} 
we present the radio source sample in the \astrobj{A 521} region;  
in Section \ref{sec:optical} we give the radio--optical identifications 
and present the nuclear radio activity of cluster early--type
galaxies; the relic radio source is presented 
in Section \ref{sec:relic}; the discussion is carried out in
Section \ref{sec:discussion}; our conclusions are summarized in
Section \ref{sec:summary}.
\\
\\
Throughout the paper we assume $H_0 =70$~km~s$^{-1}$~Mpc$^{-1}$,
$\Omega_m$=0.3 and $\Omega_{\Lambda}$=0.7. 
At the average redshift of \astrobj{A 521} ($z=0.247$), 
this cosmology leads to a linear scale of 
1 arcsec= 3.87 kpc.

\section{The cluster Abell 521}\label{sec:a521}

Abell 521 is a rich galaxy cluster, 
located at a mean redshift z=0.247. 
Its general properties are given in Table \ref{tab:prop}.
Note that the cluster coordinates in the table are those of
the ROSAT/HRI X--ray centre (A00); 
the X--ray luminosity (0.1--2 keV band) is from B\"ohringer et 
al. 2004 (REFLEX galaxy cluster catalogue); the 
velocity dispersion $\sigma_v$ is from F03; the temperature 
value is taken from A00; the virial mass M$_{\rm V}$ was 
computed from the L$_X$--M$_{\rm V}$ relation in 
Reiprich \& B\"ohriger (2002), adopting the cosmology used 
in this paper; R$_{\rm V}$ is the corresponding virial radius.

\begin{table} 
\caption[]{Properties of the cluster \astrobj{A 521}.}
\smallskip
%\begin{center}
\begin{tabular}{lc}
\hline\noalign{\smallskip}
RA$_{J2000}$       & \hspace{1mm} 04$^h$ 54$^m$ 08.6$^s$ \\
DEC$_{J2000}$      & $-$10$^{\circ}$ 14$^{\prime}$ 39.0$^{\prime \prime}$ \\
Bautz--Morgan Class & III \\
Richness            &  1  \\
Redshift            &  0.247 \\
$\sigma_v$ (km s$^{-1}$) &   1325 \\
L$_X$ (10$^{44}$ erg s$^{-1}$) &  8.178\\
T$_X$ (keV)    &  6.4 \\
M$_{\rm V}$ (M$_{\odot}$) & 1.8 $\times$ 10$^{15}$ \\
R$_{\rm V}$ (Mpc)  & 2.8 \\
\hline
\end{tabular}
%\end{center}
\label{tab:prop}
\end{table}

\begin{figure}
\centering
\includegraphics[angle=0,width=10cm]{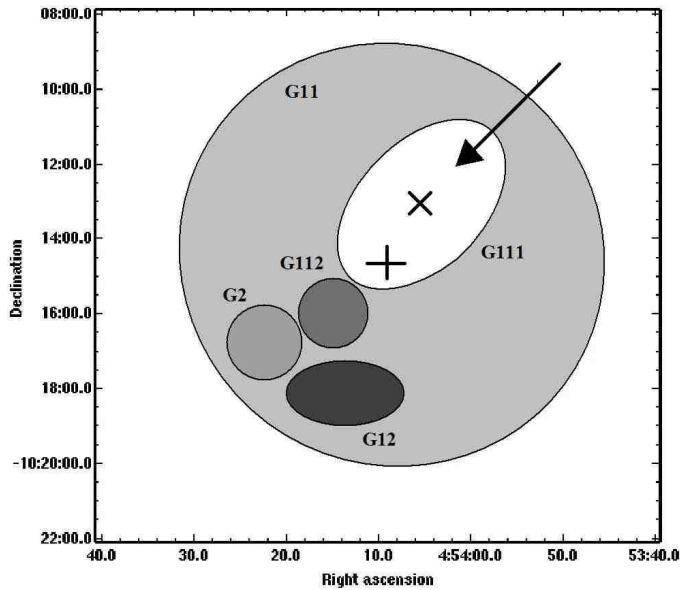}
\caption{Cartoon  representing the substructures
in \astrobj{A 521}. The cross marks the X--ray centre of the main
cluster G11, and the x--point indicates the centre of the X--ray 
compact group associated with the optical group G111 (see A00). 
The arrow represents the possible infalling direction of G111.}
\label{fig:cartoon}
\end{figure}

Detailed X--ray (A00, F05) and optical studies of \astrobj{A 521} (M00, F03)
revealed that the dynamical state of this cluster is 
very complex, since it is still undergoing multiple 
merging events. Figure \ref{fig:cartoon} sketches the scenario
proposed by these authors.

From the X--ray analysis A00 
concluded that the main merger episode
is occurring along the North--West/South--East  
direction (arrow in Fig. 1), between the main cluster (G11) and a 
northwestern compact group (G111), whose X--ray emissions 
are centered on RA=04$^h$ 54$^m$ 08.6$^s$, 
DEC=$-$10$^{\circ}$ 14$^{\prime}$ 39.0$^{\prime \prime}$
and RA=04$^h$ 54$^m$ 05.8$^s$, DEC=$-$10$^{\circ}$ 
13$^{\prime}$ 00.4$^{\prime \prime}$ respectively. 
The gas mass ratio between G11 and G111  
components is M$_{\rm gas,main}$/M$_{\rm gas, sub} \sim7$ (A00).
In their more recent analysis F05 reported a misalignment
between the  X--ray and optical merger axis.

The study of the galaxy distribution and the substructure
analysis carried out by F03 revealed the 
existence of many optical subclumps aligned along the 
merger direction. In particular these authors
identified the following velocity groups along this axis 
(see Figure \ref{fig:cartoon}):

\parn $-$ G11: the main cluster with a mean velocity 
$<$v$>=73965$ km s$^{-1}$ and a velocity dispersion 
$\sigma \sim930$ km s$^{-1}$;

\parn $-$ G111: a group dynamically bound to the brightest 
cluster galaxy (BCG), with a very low velocity dispersion 
($\sigma \sim 250$ km s$^{-1}$) and a slightly higher mean 
velocity ($<$v$>$=74290 km s$^{-1}$) as compared to the main 
cluster G11. 
This group is associated with the compact X--ray group 
detected by A00,  which is probably falling onto the cluster 
from the NW direction.
The small difference in the mean velocity between
G111 and G11 suggests that the merging is likely to take 
place  on the plane of the sky.

\parn $-$ G112: a compact group bound to the cluster, whose 
velocity ($<$v$>$=74068 km s$^{-1}$, 
$\sigma \sim 570$ km s$^{-1}$) is similar to that of the infalling 
group G111. The virial masses estimated for G111 and G112 in F03
on the basis of the optical information are much smaller 
($\sim$ one order of magnitude) than that of the main cluster G11.
We note that the mass ratios between G11 and the infalling 
groups estimated from the optical information are even larger
than those derived from the X--ray data.

\parn $-$ G12: the lowest mass group bound to \astrobj{A 521}, with
higher velocity galaxies 
($<$v$>$=75730 km s$^{-1}$, $\sigma \sim$120 km s$^{-1}$).

\parn $-$ G2: a group South--East of \astrobj{A 521}, at a
projected distance of $\sim$ 900 kpc from the X--ray centre of the main
cluster G11. This group has a mean velocity of $<$v$>$=78418 km s$^{-1}$ 
($\sigma \sim$ 500 km s$^{-1}$), which is much higher than the  cluster
velocity. On the basis of the two--body criteria, F03 concluded that this 
group is probably not bound to \astrobj{A 521}.
\\
Furthermore, F03 found also evidence of a filamentary structure
of galaxies in the central region of the cluster, extending 
along the NE-SW direction, with velocity 
$<$v$>$=73625 km s$^{-1}$ and high velocity dispersion.
This structure has been interpreted by F03 as 
evidence of an older merger, which occurred along a direction
orthogonal to the axis of the presently ongoing merger. 

\section{Radio observations}\label{sec:obs}

\begin{table} 
\caption[]{Details of the GMRT observations for \astrobj{A 521}.}
\smallskip
%\begin{center}
\begin{tabular}{lc}
\hline\noalign{\smallskip}
Observation date   & 7 -- 8 January 2005\\
$\nu$       & 610 MHz  \\
$\Delta \nu$& 32 MHz\\
RA$_{J2000}$       &  \hspace{1mm} 04$^h$ 54$^m$ 09$^s$ \\
DEC$_{J2000}$      & --10$^{\circ}$ 14$^{\prime}$ 19$^{\prime \prime}$ \\
Primary beam & 43$^{\prime}$ \\
Restoring beam, PA (full array) & 8.6$^{\prime \prime}\times$4.0$^{\prime \prime}$, 57$^{\circ}$ \\
Total Obs.time  & 3.5 h \\
rms    & 35 $\mu$Jy b$^{-1}$ \\
\hline{\smallskip}
\end{tabular}
%\end{center}
\label{tab:obs}
\end{table}

%%%%%%%%%%%%%%%%%%%%%%%%%%%%%%%%%%%%%%%%%%%%%%%%%%%Figura campo
\begin{figure}
\centering
\includegraphics[angle=0,width=14cm]{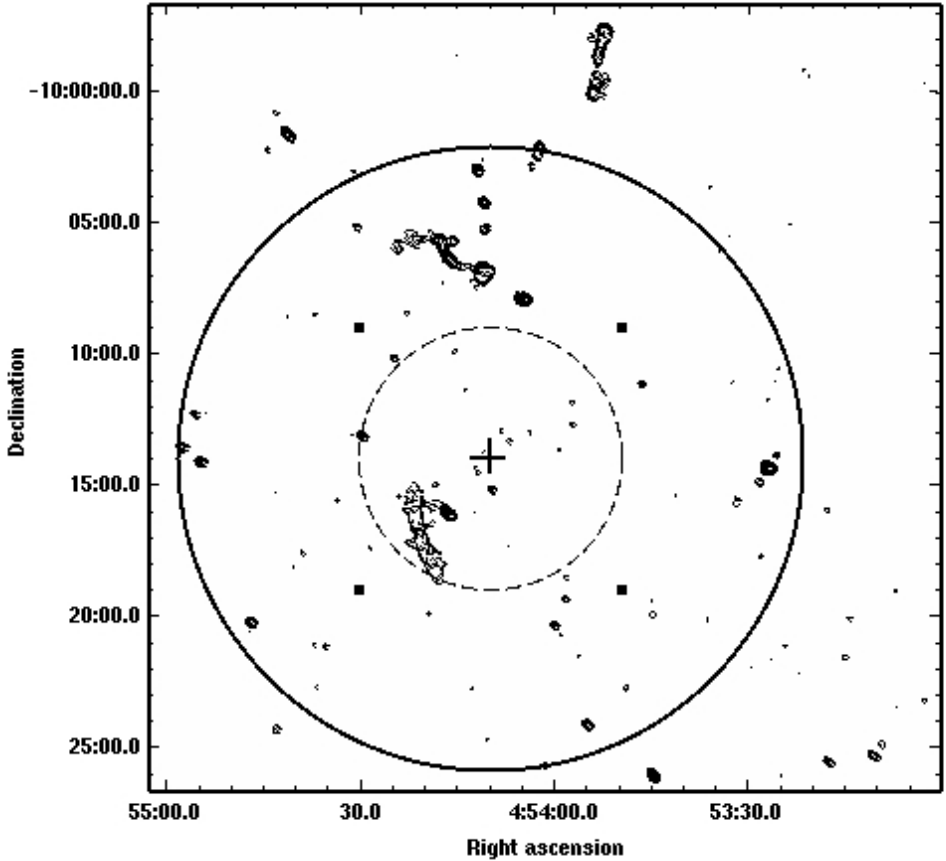}
\caption{610 MHz radio contours of the 30$^{\prime} \times
30^{\prime}$ region containing \astrobj{A 521}. 
The contours are logarithmic, and the lowest level is the 
5$\sigma$=0.2 mJy b$^{-1}$.
The resolution is $13.1^{\prime \prime} 
\times 8.1^{\prime \prime}$, in p.a. $56^{\circ}$.
The cross marks the X-ray centre of the cluster. The solid circle
has a radius corresponding to the virial radius of \astrobj{A 521}. The dashed 
circle indicates the region with the redshift information, and 
corresponds to the region shown in Fig. 3.}
\label{fig:field}
\end{figure}

We observed the cluster \astrobj{A 521} 
at 610 MHz with the GMRT on 7 and 8 January 2005,
using simultaneously the R (USB) and L (LSB) bands of 16 
MHz each, for a total frequency band of 32 MHz. 
Table \ref{tab:obs} gives the details of the observations.

The observations were carried out in spectral line mode, with 
128 channels in each band, with a spectral resolution of 125 kHz.

The four data sets (7 and 8 Jan 2005, USB and LSB) were analysed 
individually. The data calibration and reduction were performed using 
the NRAO Astronomical Image Processing System (AIPS) 
package. The sources 3C\,147 and 3C\,48 were observed as primary
calibrators at the beginning and at the end of the observations, 
to determine and correct for the bandpass shape and 
for the initial amplitude and phase calibration. The source 
0447--220 was used as secondary phase calibrator and was 
observed every 20 minutes throughout the observation. 
In order to reduce the size of the data set, after the bandpass calibration
the central channels of each data set were averaged to 6 channels of 
$\sim 2$ MHz each.

For each data set, images were produced using the 
wide--field imaging technique, with 25 facets covering 
a total field of view of $\sim 1\times1$ square degree.
After a number of phase self--calibration cycles, 
a final step was made, allowing for phase and amplitude 
corrections in  order to improve the quality of the final images.
The residual errors in the estimated flux density are $\ltsim$ 5$\%$.

The four self--calibrated data sets were then averaged from 6 channels 
to 1 single channel and finally combined together using the AIPS task DBCON. 
We note that bandwidth smearing is relevant only at the
edge of the wide field, and does not affect the region presented 
and analysed here, i.e. the inner  30$^{\prime}\times30^{\prime}$
(see Section \ref{sec:sample}).
The images from the combined data set were obtained using again 
the wide--field imaging technique, combined with the task FLATN and 
finally corrected for the primary beam appropriate to the GMRT antennas
at 610 MHz.  

We produced images over a range of resolutions, reaching rms noise 
levels of the order of $1\sigma \sim 35-40$ $\mu$Jy b$^{-1}$. 
For the purpose of the present paper we show only the image tapered to a 
resolution of 13.1$^{\prime \prime}\times$8.1$^{\prime \prime}$, in p.a. 56$^{\circ}$.
The 5$\sigma$ detection limit in this image, 0.20 mJy b$^{-1}$, corresponds
to a radio power limit of 3.5$\times 10^{22}$ W Hz$^{-1}$.

\section{The sample of radio sources}\label{sec:sample}

The total field of view of our observations 
is $\sim$ 1$^{\circ}$ $\times$ 1$^{\circ}$, which is 
much larger than the cluster size.
Here we present only a portion of the whole field, 
with size of 30$^{\prime}\times30^{\prime}$, centered on
RA=04$^{h}$ 53$^{m}$ 00$^{s}\div04^{h}$ 55$^{m}$ 00$^{s}$
and DEC=$-09^{\circ}$ 55$^{\prime}$ 00$^{\prime \prime}\div
-10^{\circ}$ 25$^{\prime}$ 00$^{\prime \prime}$. 
At the cluster distance this corresponds to a region as
large as $\sim 7\times7$ Mpc$^2$.

The 610 MHz radio emission from this region 
is shown in Figure \ref{fig:field}. In the figure we also plotted 
a solid circle with a radius corresponding to the virial radius of \astrobj{A 521} (Tab. 1),
and a dashed circle, representing the 5' radius region covered by the redshift
catalogues in M00 and F03. The cross marks the X-ray centre of the cluster
(Tab. 1).

The radio emission from the \astrobj{A 521} region is dominated by point--like 
and marginally resolved sources. However, the whole field is 
characterised by the presence of three extended radio sources. 
Two of them (see Section  \ref{sec:extended}) are located in the northern 
part of the field and are associated with early--type galaxies without
redshift information.
The elongated structure located South--East with respect to the cluster 
centre is the relic source discussed in Section \ref{sec:relic}.

We used the AIPS task SAD to identify sources in the final 13.1$^{\prime \prime}$
 $\times$ 8.1$^{\prime \prime}$ image of \astrobj{A 521} (1$\sigma$= 40 $\mu$Jy b$^{-1}$).
Given a radio image, this task (1) finds all potential sources whose peak is 
brighter than a given level; (2) Gaussian components are fitted; (3) positions 
and flux density values are given. The task also produces a residual image, 
which can be inspected to identify both extended sources not well fitted 
by Gaussians, and sources with peak flux density lower than the previous 
threshold.
As first step, we used SAD to find all sources with peak flux density
greater than 0.32 mJy b$^{-1}$, i.e. 8 times the rms noise level in 
the field. Then, on the residual image we searched for all sources with 
peak flux densities in the range 5$\sigma$ -- 8$\sigma$ (i.e. 0.20 -- 0.32
mJy b$^{-1}$). On this image we identified also the extended sources.
\\
Each radio source of the list was then carefully inspected, and 
the flux density values given by SAD for the unresolved or 
marginally resolved sources were checked using the task JMFIT. 
For the extended sources the flux density was obtained by means of TVSTAT.

The final list of radio sources (over the whole
$\sim$ 1$^{\circ}$ $\times$ 1$^{\circ}$ field), contains a total of 
101 radio sources above the peak flux density limit of 0.20 mJy b$^{-1}$; 
52 radio sources out of the total are located in the 
30$^{\prime}\times$30$^{\prime}$ region shown in Figure \ref{fig:field}, 
and are presented in Table 3, where we give:

\parn $-$ columns 1, 2 and 3: respectively name (GMRT--) and J2000 position;

\parn $-$ column 4 and 5: respectively peak and integrated flux density at 
610 MHz, corrected for the primary beam attenuation.
Note that the flux density given for the relic source 
(J0454--1017a) does not include the embedded point sources, whose
flux density was estimated from the full resolution image 
($8.6^{\prime\prime} \times 4.0^{\prime\prime}$, see Table 1) 
and subtracted;

\parn $-$ column 6: radio morphology. We classified the sources as unres.= 
unresolved and ext. = extended. Moreover we indicated as WAT a 
wide--angle--tailed morphology, D a double structure and Rel the relic source.
For the double sources we give the position of the radio 
barycentre and for the extended sources we give the position of the radio peak.

%
%------------- Table 3-------------------------------------------------------------
%
\begin{table*}
\label{tab:cat}
\caption[]{Source list and flux density values}
\begin{center}
\begin{tabular}{lccccc}
\hline\noalign{\smallskip}
Name   & RA$_{J2000}$ &  DEC$_{J2000}$ & S$_{\rm peak}$ & S$_{\rm tot}$ & Radio Morphology \\ 
GMRT$-$&              &                & mJy b$^{-1}$ &  mJy    &                  \\ 
\noalign{\smallskip}
\hline\noalign{\smallskip}
J0453$-$1023  &  04 53 02.69 & $-$10 23 12.0 &  0.74 &    0.89 & unres.\\
J0453$-$1020a &  04 53 14.19 & $-$10 20 05.4 &  0.54 &    0.67 & unres.\\
J0453$-$1021  &  04 53 14.97 & $-$10 21 33.0 &  0.68 &    0.91 & unres.\\
J0453$-$1015  &  04 53 17.70 & $-$10 15 58.0 &  0.53 &    0.59 & unres.\\
J0453$-$1014  &  04 53 26.84 & $-$10 14 19.7 & 74.43 &   96.56 & unres.\\
J0453$-$1019a &  04 53 44.72 & $-$10 19 55.2 &  0.48 &    0.78 & unres.\\
J0453$-$1011a &  04 53 46.43 & $-$10 11 09.3 &  0.72 &    0.79 & unres.\\
J0453$-$1022  &  04 53 48.85 & $-$10 22 43.5 &  0.45 &    0.64 & unres.\\
J0453$-$0957  &  04 53 52.14 & $-$09 57 59.0 &  4.84 &   49.73 & D\\
J0453$-$1024  &  04 53 54.80 & $-$10 24 08.2 &  3.72 &    4.59 & unres.\\
J0453$-$1012  &  04 53 57.05 & $-$10 12 41.7 &  0.51 &    0.54 & unres.\\
J0453$-$1011b &  04 53 57.23 & $-$10 11 51.2 &  0.35 &    0.52 & unres.\\
J0453$-$1018  &  04 53 58.04 & $-$10 18 32.1 &  0.42 &    0.43 & unres.\\
J0453$-$1019b &  04 53 58.21 & $-$10 19 21.4 &  0.89 &    0.90 & unres.\\
J0453$-$1020b &  04 53 59.87 & $-$10 20 21.1 &  1.47 &    1.54 & unres.\\
J0454$-$1002a &  04 54 02.26 & $-$10 02 07.0 &  3.68 &    9.11 & ext.\\
J0454$-$1002b &  04 54 03.45 & $-$10 02 50.1 &  0.73 &    1.02 & unres.\\
J0454$-$1013a &  04 54 03.60 & $-$10 13 00.1 &  0.33 &    0.33 & unres.\\
J0454$-$1007  &  04 54 04.40 & $-$10 07 53.5 & 16.43 &   36.77 & D     \\
J0454$-$1013b &  04 54 06.90 & $-$10 13 20.7 &  0.32 &    0.47 & unres.\\
J0454$-$1012a &  04 54 08.15 & $-$10 12 55.0 &  0.31 &    0.46 & unres.\\
J0454$-$1015  &  04 54 09.53 & $-$10 15 09.6 &  1.47 &    1.49 & unres.\\
J0454$-$1005a &  04 54 10.79 & $-$10 05 16.0 &  1.21 &    2.33 & ext.\\
J0454$-$1004  &  04 54 10.82 & $-$10 04 14.1 &  5.99 &    7.69 & unres.\\
J0454$-$1013c &  04 54 10.94 & $-$10 13 43.6 &  0.27 &    0.33 & unres. \\ 
J0454$-$1014a &  04 54 11.66 & $-$10 14 33.2 &  0.30 &    0.47 & unres. \\
J0454$-$1002c &  04 54 11.82 & $-$10 02 59.00 &  5.37 &    7.46 & unres.\\
J0454$-$1014b &  04 54 12.08 & $-$10 14 19.6 &  0.33 &    0.42 & unres. \\
J0454$-$1011  &  04 54 13.71 & $-$10 11 21.9 &  0.30 &    0.37 & unres. \\
\noalign{\smallskip}
\hline
\end{tabular}
\end{center}
\end{table*}

\setcounter{table}{2}
\begin{table*}
\label{tab:cat}
\caption[]{Continued}
\begin{center}
\begin{tabular}{lccccc}
\noalign{\smallskip}
\hline\noalign{\smallskip}
J0454$-$1013d &  04 54 13.93 & $-$10 13 31.4 &  0.24 &    0.28 & unres. \\ 
J0454$-$1009  &  04 54 15.39 & $-$10 09 53.8 &  0.47 &    0.47 & unres.\\
J0454$-$1005b &  04 54 15.68 & $-$10 05 42.8 &  1.84 &    2.89 & unres.\\
J0454$-$1016a &  04 54 16.28 & $-$10 16 05.9 & 22.93 &   27.56 & unres.\\
J0454$-$1006  &  04 54 16.33 & $-$10 06 16.4 & 23.84 &  104.36 & WAT\\
J0454$-$1019  &  04 54 19.34 & $-$10 19 53.2 &  0.37 &    0.38 & unres.\\
J0454$-$1017a  &   04 54 20  & $-$10 17 00  &   $-$      &   41.9  & Rel. \\
J0454$-$1016b &  04 54 21.02 & $-$10 16 44.2 &  1.77 &    3.85 & unres.\\
J0454$-$1017b  &  04 54 22.34 & $-$10 17 11.9 &  0.71 &    1.13 & unres.\\
J0454$-$1008a &  04 54 22.79 & $-$10 08 26.9 &  0.38 &    0.38 & unres.\\
J0454$-$1010  &  04 54 24.73 & $-$10 10 09.4 &  1.01 &    1.04 & unres.\\
J0454$-$1013c &  04 54 29.60 & $-$10 13 08.2 &  4.48 &    4.55 & unres.\\
J0454$-$1005c &  04 54 30.41 & $-$10 05 09.8 &  1.14 &    1.27 & unres.\\
J0454$-$1021a &  04 54 35.33 & $-$10 21 10.0 &  0.59 &    0.67 & unres.\\
J0454$-$1008b &  04 54 36.97 & $-$10 08 30.0 &  0.39 &    0.40 & unres.\\
J0454$-$1001  &  04 54 41.14 & $-$10 01 37.6 & 16.54 &   24.35 & unres.\\
J0454$-$1024  &  04 54 42.89 & $-$10 24 17.5 &  0.70 &    0.93 & ext.\\
J0454$-$1002d &  04 54 44.20 & $-$10 02 11.6 &  0.75 &    0.95 & unres.\\
J0454$-$1020  &  04 54 46.79 & $-$10 20 14.5 &  4.31 &    5.68 & unres.\\
J0454$-$1014c &  04 54 54.60 & $-$10 14 06.5 &  5.42 &    6.82 & unres.\\
J0454$-$1012b &  04 54 55.54 & $-$10 12 16.7 &  1.44 &    1.98 & unres.\\
J0454$-$0955  &  04 54 57.27 & $-$09 55 41.5 &  1.64 &    2.46 & ext.\\
J0454$-$1013e &  04 54 57.55 & $-$10 13 33.0 &  3.82 &    5.24 & unres.\\
\noalign{\smallskip}
\hline
\end{tabular}
\end{center}
\end{table*}
%
%------------------- end of table 3------------------------------------------------------
%

We detected all the 20 radio sources found by A00 in their analysis of a portion
of the 1.4 GHz NRAO VLA Sky Survey (NVSS, Condon et al. 1996) 
image, with a size similar to the field shown in Figure \ref{fig:field}.
We note that their sources labelled 12, 14 and 15 are part of the diffuse 
radio relic (see Section \ref{sec:relic}), and their sources labelled 7, 11 
and 13 are part of the wide--angle tail described in Section 5.2.
The remaining radio sources listed in Table 3 are either undetected or
only marginally visible in the NVSS, due to the different resolution and the 
sensitivity limit of the NVSS survey (1$\sigma$=0.45 mJy b$^{-1}$).

\section{Optical Identifications}\label{sec:optical}

The sample of 52 radio sources presented in the previous section 
was cross--correlated with the Super COSMOS/UKST 
Southern Sky Object Catalogue (Hambly et al. 2001) 
and the APM Catalogue (Maddox et al. 1990), 
to search for optical counterparts.
Radio/optical overlays (using the DSS--1) were visually inspected 
for all the candidate identifications, and for the remaining radio 
sources in the sample, in order to find possible optical counterparts 
lost by the incompleteness of 
these catalogues. 

We estimated the reliability of the optical identifications on the
basis of the parameter $R$, which takes into account the
uncertainty in the radio and optical positions:

$$ R^2 = \frac{\Delta^2_{r-o}}{\sigma^2_o + \sigma^2_r} $$

\noindent where $\Delta_{r-o}$ is the offset between the radio 
and optical coordinates, and $\sigma_o$ and $\sigma_r$ 
are the optical and the radio position errors respectively.
We adopted a mean positional uncertainty of $\sigma_o$=1.5 arcsec
for the optical catalogues (Unewisse et al. 1993), 
and with the parameters of our
observations we estimated an average radio positional error 
of 1 arcsec both in right ascension and declination
(Prandoni et al. 2000).

We considered reliable identifications all matches with 
$R \le 3$, i.e. we assume that
all matches $\le 3$ are due to the random distribution 
of the positional errors, while for R $>$ 3 the difference 
between the optical and radio position is significant.
For two sources we found $R > 3$,
and therefore we cosidered them uncertain identification
(see notes to Table 4).
We found 21 radio--optical identifications (including
the 2 uncertain cases), which correspond 
to $40\%$ of our radio source sample. 

In order to find
the sources associated with \astrobj{A 521} member galaxies, 
we cross--correlated our sample of identified sources 
with the redshift catalogues in M00 and F03. 
We note that these two catalogues do not cover the full
region of $30^{\prime} \times 30^{\prime}$ analysed in the
present paper, therefore our redshift search is actually 
restricted to a region of $\sim 5^{\prime}$ (1.1 Mpc) 
radius from the cluster centre (Fig. \ref{fig:field}). 
This region includes 17 radio sources,
11 of which with an optical counterpart. 
Among these, 8 radio galaxies are located within the
velocity range 70000 -- 80000 km/s. One of them, however
(J0454--1016b) is located in the group G2 (see Sect. 2), 
considered to be unbound in F03, therefore it will not be
considered in the analysis presented in Sect. 5.3.

The list of the radio--optical identifications is reported in 
Table 4, where we give:

\parn $-$ column 1: radio and optical name, where FMC and MBP stands for 
optical counterparts from F03 and M00 respectively;

\parn $-$ columns 2 and 3: J2000 radio and optical coordinates;

\parn $-$ column 4: integrated flux density at 610 MHz and $I$ magnitude
given by the SuperCOSMOS or the APM catalogue when available,
otherwise determined from the R magnitude adopting the  
(R--I)=0.77 colour for early--type galaxies at z=0.2 (Fukugita et al. 
1995). The I magnitudes are corrected
for a galactic absorption of A$_I$=0.146 (Schlegel et al. 1998); 
if they were derived from the R magnitudes, we first corrected 
these latter using an absorption A$_R$=0.201;

\parn $-$ column 5: radio morphology and (R--I) colour from the
SuperCOSMOS or the APM catalogue;

\parn $-$ column 6: radio power at 610 MHz and radial velocity;

\parn $-$ column 7: information about the galaxy type, and
spectral features of the optical counterpart from
M00 and F03 and $R$ parameter. Note that for the
extended radio galaxies discussed in Section 5.2, 
no value of $R$ is given, due to the extended and complex
radio morphology.

%
%---------------- Table 4-----------------------------------------------------
%
\setcounter{table}{3}
\begin{table*}
\label{tab:optic}
\caption[]{Optical Identifications}
\begin{center}
\scriptsize
\begin{tabular}{lllcccc}
\hline\noalign{\smallskip}
Radio Name    & RA$_{J2000}$  & DEC$_{J2000}$ & $S_{610 \rm MHz}$ & Radio type & logP$_{610 \rm MHz}$ & optical notes\\
GMRT$-$        &               &               &  (mJy)       &            & (W Hz$^{-1}$)  &\\
Opt. Name     & RA$_{J2000}$  & DEC$_{J2000}$ &   I    &  (R--I) & v& $R$         \\
              &               &               &              &         &    (km s$^{-1}$) &   \\
\noalign{\smallskip}
\hline\noalign{\smallskip}
J0453$-$1023  & 04 53 02.69  & $-$10 23 12.0 & 0.89 & unres. & $-$ &  \\
              & 04 53 03.31  & $-$10 23 06.2 & 16.20 & 0.93 & $-$ & 6.0  (*)\\
&&&&&&\\
J0453$-$1015  & 04 53 17.70  & $-$10 15 58.0 & 0.59& unres. & $-$ &\\
              & 04 53 17.62  & $-$10 16 00.6  & 18.59 & 0.17& $-$ & 1.6 \\
&&&&&&\\
J0453$-$1019a &  04 53 44.72 & $-$10 19 55.2 &  0.77 &  unres. & $-$ &    \\
               & 04 53 44.45&  $-$10 19 58.7 & 15.58 & 0.93 & $-$ & 2.9  \\
&&&&& &\\
J0453$-$0957  & 04 53 52.14  &$-$09 57 59.0  & 49.73  & D  &$-$& \\
            & 04 53 53.07 & $-$09 58 35.4 & 16.76 & 1.00 & $-$ & $-$ \\
&&&&&& \\
J0453$-$1024  & 04 53 54.80  & $-$10 24 08.2 & 4.59  &  unres. & $-$ &  \\
               & 04 53 55.25  & $-$10 24 15.4 & $-$ &$-$ & $-$ & 5.3 (**) \\
&&&&&&\\
J0453$-$1012  &  04 53 57.05 & $-$10 12 41.7 &  0.54 & unres.  &   22.96 & early--type  \\
MBP20   & 04 53 57.00& $-$10 12 44.9 &  16.67  & 0.94 & 73044  & 1.8 \\
&&&&&& \\
J0453$-$1019b &  04 53 58.21 & $-$10 19 21.4 & 0.90 & unres. & $-$ &   \\ 
     &  04 53 58.26 & $-$10 19 21.0 & 18.67     &   $-$   &$-$ &  0.4      \\
&&&&&& \\
J0454$-$1013a & 04 54 03.60 & $-$10 13 00.1 &  0.33 & unres.   & 22.77 & early--type     \\
MBP14   & 04 54 03.58 & $-$10 12 59.6 &  17.90  & 0.92 &   75146 & 0.3 \\
&&&&& &\\
J0454$-$1013b & 04 54 06.90 & $-$10 13 20.7 &  0.47 &  unres.  & 22.92 &  early--type, BCG    \\
FMC65                  & 04 54 06.84 & $-$10 13 23.3 &  16.19 &1.10 & 74372  & 1.5  \\ 
&&&&&&\\
J0454$-$1012a &  04 54 08.15 & $-$10 12 55.0 & 0.46 & unres.  &  22.92 &  early--type   \\
MBP11   & 04 54 08.11 & $-$10 12 54.0  &  19.00  & 0.92&      74873  & 0.6 \\
 &&&&&& \\      
J0454$-$1015 & 04 54 09.53 & $-$10 15 09.6 &  1.49 & unres.   &    22.85   \\
FMC90  & 04 54 09.33 & $-$10 15 10.1 &  17.74       &$-$  &  40747 & 1.7 \\
&&&&&& \\
J0454$-$1014a & 04 54 11.66 & $-$10 14 33.2 & 0.47 & unres. &  22.87 &  late--type, OII,OIIIa,b,H$\beta$\\
FMC105                            & 04 54 11.72 & $-$10 14 35.3 &   18.93 & 0.63 & 70952  &1.2 \\
&&&&&& \\
J0454$-$1013d& 04 54 13.93 & $-$10 13 31.3 & 0.28 & unres. &   22.86     &  OII, Balmer\\
FMC121                            &  04 54 13.90 & $-$10 13 32.3 &  18.12    &  0.79 & 88714  & 0.6\\
\noalign{\smallskip}
\hline\noalign{\smallskip}
\noalign{\smallskip}
\end{tabular}
\end{center}
\end{table*}

\setcounter{table}{3}
\begin{table*}
\label{tab:optic}
\caption[]{Continued}
\begin{center}
\scriptsize
\begin{tabular}{lllcccc}
\noalign{\smallskip}
\hline\noalign{\smallskip}
 J0454$-$1016a & 04 54 16.28 & $-$10 16 05.9 & 27.56 & unres. &   24.69 &  \\
FMC143 & 04 54 16.34 & $-$10 16 04.6  & 17.00 &  1.26  & 74282  & 0.9\\
&&&&&& \\
J0454$-$1006   & 05 54 16.28& $-$10 06 16.4 & 103.36 & WAT  &$-$ & \\
               & 05 54 16.05 & $-$10 06 24.6 & 16.69 & 0.91 & $-$ &$-$  \\
&&&&&&\\
J0454$-$1016b & 04 54 21.02 & $-$10 16 44.2 &  3.85 & unres. & 23.88 &   late--type, OII, H$\beta$\\
FMC170  & 04 54 20.96 & $-$10 16 44.8 & 18.35 &  0.88  & 78326 & 0.5\\
&&&&&& \\
J0454$-$1017b &  04 54 22.34 & $-$10 17 11.9 & 1.13 & unres. &   23.28 &  late--type,  H$\beta$\\
FMC178 & 04 54 22.35 & $-$10 17 13.4 & 17.20 &  0.75  &  72593   & 0.8\\
 &&&&& \\
J0454$-$1008a &  04 54 22.79 & $-$10 08 26.9 &  0.38 &  unres. & $-$     & \\
                        & 04 54 22.97 & $-$10 08 25.7 & 18.05  & $-$    & $-$   & 1.6\\
 &&&&&& \\     
J0454$-$1010 & 04 54 24.73 & $-$10 10 09.4 &  1.04 & unres.   &  $-$   &  \\ 
              & 04 54 24.65 & $-$10 10 10.0 & 17.47&  0.85  &   $-$  & 0.8 \\    
&&&&&& \\
J0454$-$1005c  & 04 54 30.41& $-$10 05 09.8 & 1.27   &  unres.    & $-$& \\
               & 04 54 30.25 & $-$10 05 11.9 & 16.75 & 0.92  &$-$ &1.7 \\
&&&&&&\\
 J0454$-$1008b &  04 54 36.97 & $-$10 08 30.0 & 0.39 & unres.   & $-$   &   \\
 & 04 54 36.72  & $-$10 08 32.7 & 18.19  &  0.28 & $-$ & 2.5\\
\noalign{\smallskip}
\hline\noalign{\smallskip}
\noalign{\smallskip}
\end{tabular}
\end{center}
(*): this identification is uncertain since the candidate optical counterpart is
misplaced with respect to the radio emission peak;
(**): for the optical counterpart of this source only the magnitude b$_J$=21.71
is available. This galaxy falls within the radio contours of the source, but it is dislocated with 
respect to the emission peak.
\end{table*}
%
%------------------------- end of Table 4 --------------------------- 

\subsection{Unresolved radio galaxies in the \astrobj{A 521} region}\label{sec:pop}

Among the 8 cluster radio galaxies (see Table 4),
we found three late--type and four early--type galaxies. One of them,
J0454--1013b, is associated with the Brightest Cluster Galaxy (FMC65). 
For the remaining source, J0454$-$1016a, no colour information is available, 
however on the basis of its featureless spectrum (F03),
and of the (R--I)=1.26 colour (taken from the SuperCOSMOS catalogue),
we can {\it bona fide} classify it as an early--type galaxy.

\begin{figure}
\centering
\includegraphics[angle=0,width=9cm]{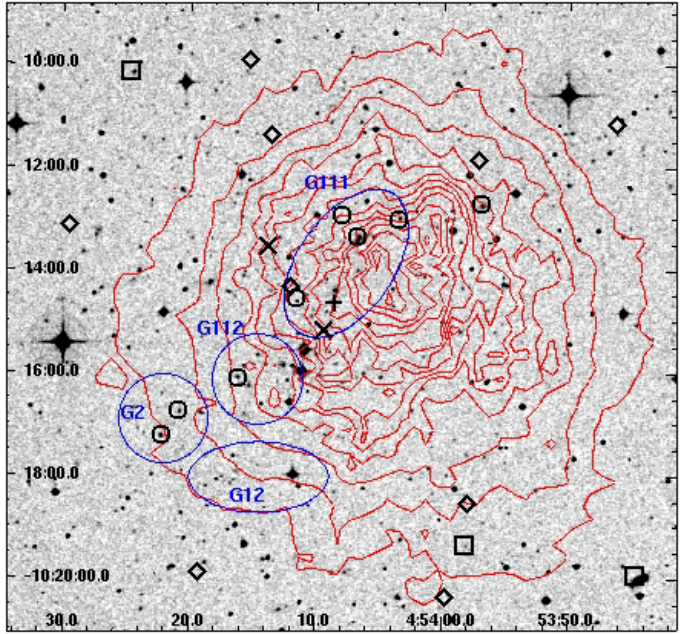}
\caption{Location of the radio sources within 1.5 Mpc from the
\astrobj{A 521} centre, overlaid on the DSS--1 optical frame image and the
X--ray ASCA contours. The cross indicates the ROSAT/HRI centre
of the cluster (Tab.1). Circles represent the 8 radio galaxies belonging to \astrobj{A 521}; 
the x--points represent the position of two radio galaxy located at
a redshift different from \astrobj{A 521}; squares are radio sources 
identified with an optical object without redshift information; 
diamonds are radio sources with no identified optical counterpart.
Large blue circles and ellipses indicate the dynamical groups of
optical galaxies. The X--ray contours are 
3.0$\times$10$^{-5} \div 2.4 \times 10^{-4}$ cts sec$^{-1}$ and
are spaced of 1.5 $\times 10^{-5}$ cts sec$^{-1}$.}
\label{fig:location}
\end{figure}

For the cluster radio galaxies we searched for segregation effects 
both in the plane of the sky and in the velocity space.
\\
In Figure \ref{fig:location} we show the distribution
of the radio galaxies within 1.5 Mpc from the cluster centre,
overlaid on the DSS--1 optical image and the X--ray 
isophotes from an archival ASCA observation ($\sim$ 45 ks). 
This X--ray image was extensively analysed in A00; 
the purpose of its use here is to illustrate 
the relative distribution of the hot gas and 
the positions of the cluster radio galaxies.
The 8 radio galaxies of \astrobj{A 521} are represented by circles, while
the two radio galaxies at a redshift different from \astrobj{A 521}
are indicated by x--points. Squares show the location of radio
sources identified with objects without redshift
information, and diamonds are radio sources with no optical
identification. Moreover the cross indicates the X--ray (ROSAT/HRI)
centre of the cluster (Tab. 1) and the blue large circles and ellipses 
represent the dynamical groups of optical galaxies 
found by F03 ( Section 2, Fig. 1). The misplacement between
the ROSAT cluster centre and the peaks of the ASCA isophotes
can be explained as due to the lower resolution of the ASCA image,
combined with the large uncertainties in the positional accuracy 
(A00).

It is clear that the radio galaxy distribution in \astrobj{A 521} is not
random. In particular, all the radio galaxies within the
cluster X--ray emission are aligned along 
the NW--SE axis, which is the suggested direction 
of the ongoing merger (A00, F05). On the other hand, no radio galaxy 
was detected along the old merger axis (roughly perpendicular to 
the NW--SE axis), where F03 found a ridge of high optical density. 
We note that a similar situation (at least in projection onto the 
plane of the sky) is found in \astrobj{A 2255} (Miller \& Owen 2003), 
where the radio galaxies are distributed along the merger axis. 

\begin{figure}
\centering
\includegraphics[angle=0,width=9cm]{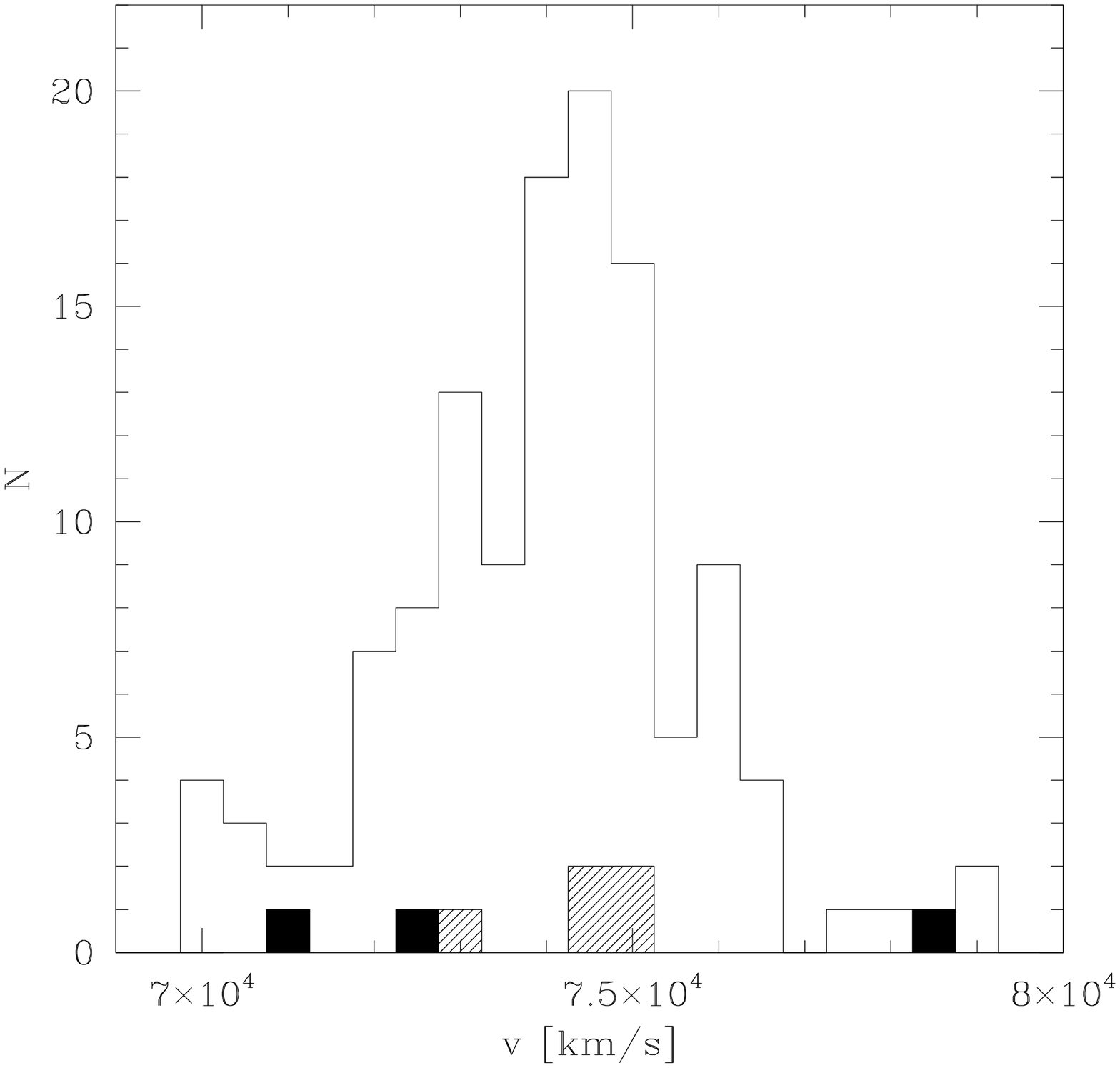}
\caption{Velocity distribution of the 8 radio galaxies of \astrobj{A 521}
compared to the redshift distribution of cluster galaxies. 
Black bins represent late--type galaxies, dashed bins 
indicate early--type galaxies. The five galaxies with the 
highest velocity belong to the unbound group G2.}
\label{fig:vel}
\end{figure}

In Figure \ref{fig:vel} the velocity distribution of
the radio galaxies in \astrobj{A 521} is compared to the 
distribution of the whole velocity sample in M00 
and F03 with secure velocity measurement. 
Objects corresponding to late--type radio galaxies are marked as 
black bins, the early--type radio galaxies are indicated by the dashed bin.
The early--type objects are all found in the bins containing
the bulk of the cluster galaxies, while the late--type
radio galaxies are at the edge of the velocity distribution,
even excluding the G2 member from these qualitative considerations.
In particular, the three radio galaxies in the northern part 
of the G111 group (the BCG group, associated with the  
group detected in the X--ray) and the source belonging to 
G112 (see Figure \ref{fig:location}) are all early--type galaxies, 
and are within a narrow range in velocity, i.e. 74282$\div$75146 km s$^{-1}$.
F03 already noted that G111 and G112 are essentially at 
the same velocity within $\sim$ 200 km s$^{-1}$ (Sect. 2).
\\
The remaining early--type radio galaxy of \astrobj{A 521} is the 
westernmost optical identification in Figure \ref{fig:location}
and has a lower velocity (73044 km s$^{-1}$), consistent 
in redshift with the main cluster G11 (Fig. 1).


Among the late--type radio galaxies, one is located 
within the G111 group boundary (see Figure \ref{fig:location}), but
has a significantly lower velocity with respect to the group.
The remaining two sources are embedded
in the relic emission (Section \ref{sec:relic}) and projected 
within the G2 group. Of these two,  
J0454$-$1016b (the northwestern one) has a velocity
consistent with the group, which is probably unbound to \astrobj{A 521}, 
while the velocity of the other is significantly lower and 
consistent with the main cluster.

\begin{figure}
\centering
\includegraphics[angle=0,width=9cm]{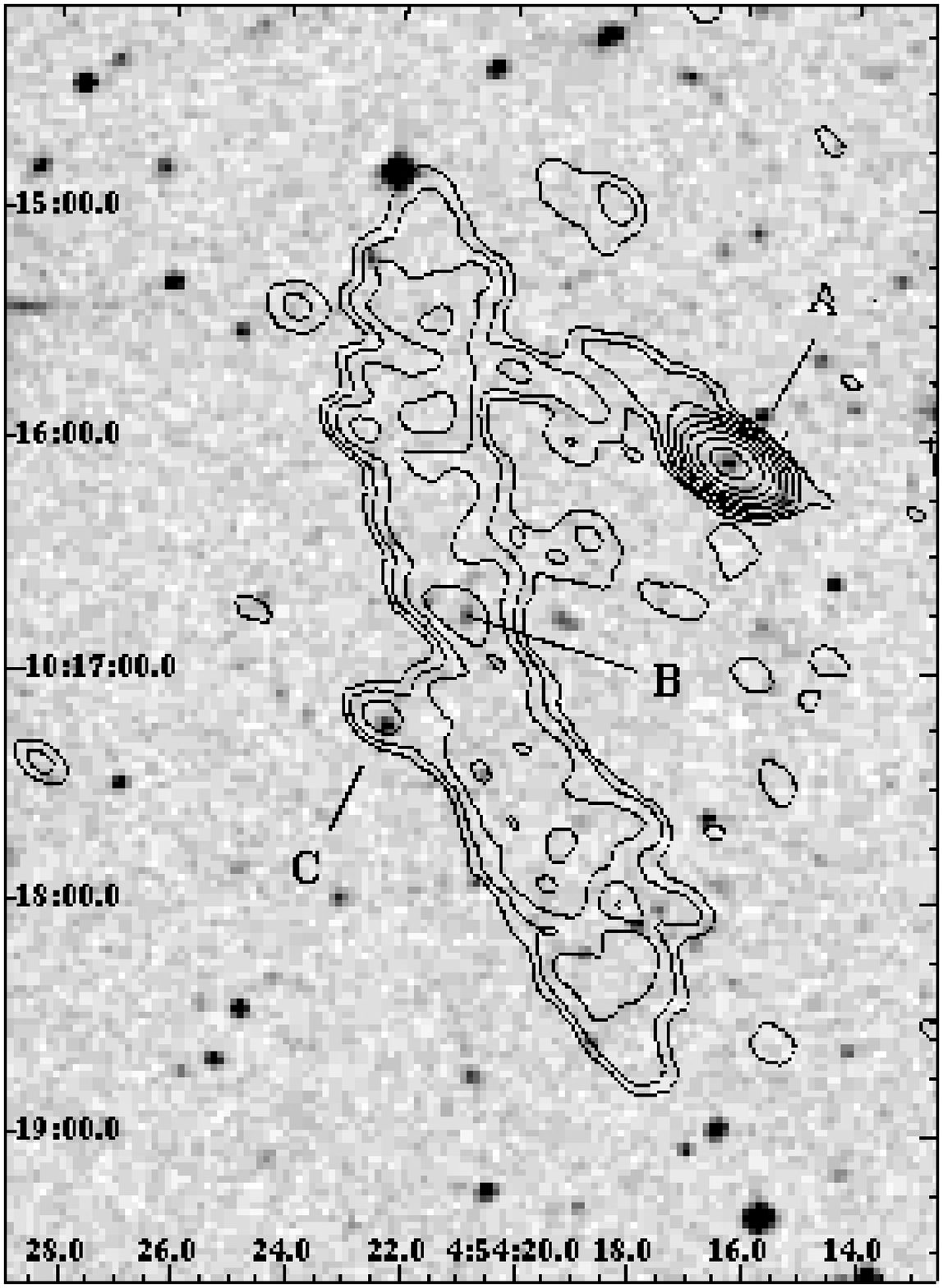}
\caption{610 MHz contours of the diffuse radio source in \astrobj{A 521},
 overlaid on the DSS--1 optical frame.
The radio contours
are 0.12 $\times$(-1, 1, 2, 4, 8, 16, 32, 64, 128, 
256, 512) mJy b$^{-1}$. The resolution is $13.1^{\prime \prime} 
\times 8.1^{\prime \prime}$, in p.a. $56^{\circ}$.
A, B and C indicates the position of radio galaxies embedded in 
the relic emission, i.e. the sources J0454--1016a,
J0454--1016b and J0454--1017 respectively.}
\label{fig:relic2}
\end{figure}

All the radio galaxies in \astrobj{A 521} have low radio power, exception
made for J0454--1016a, located in projection close to the radio relic 
(see Table 4 and
Figure \ref{fig:relic2}), with logP$_{610 \rm MHz}$ (W Hz$^{-1}$)  = 24.69. 
Interestingly, two late--type radio galaxies are more powerful than 
the early--type ones, suggesting enhanced star formation.
Unfortunately, the data available in the literature do not allow to
confirm this possibility.

\subsection{Extended radio galaxies in the \astrobj{A 521} region}\label{sec:extended}

\begin{figure}
\centering
\includegraphics[angle=0,width=11cm]{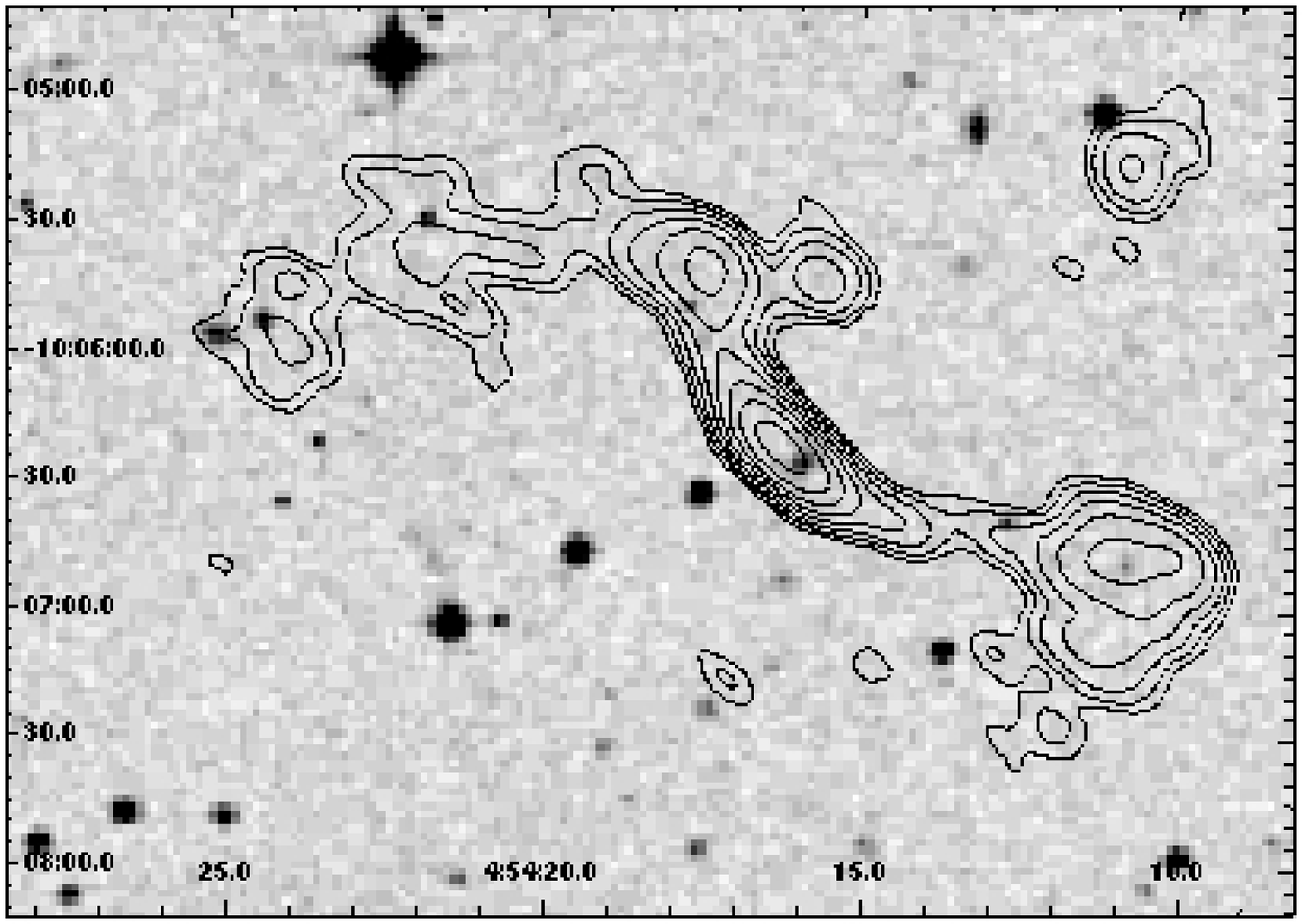}
\caption{610 MHz contours of the wide--angle--tail radio 
source J0454$-$1006, located at the North 
of \astrobj{A 521}, overlaid on the DSS--1 optical image. 
The radio contours are 0.12 $\times$(-1, 1, 2, 4, 8, 16, 
32, 64, 128, 256, 512) mJy b$^{-1}$.
 The resolution is $13.1^{\prime \prime} 
\times 8.1^{\prime \prime}$, in p.a. $56^{\circ}$. J0454$--$1005b
is a point--like source, which is given as a separate source in Table \ref{tab:cat}.}
\label{fig:WAT}
\end{figure}

\begin{figure}
\centering
\includegraphics[angle=0,width=8cm]{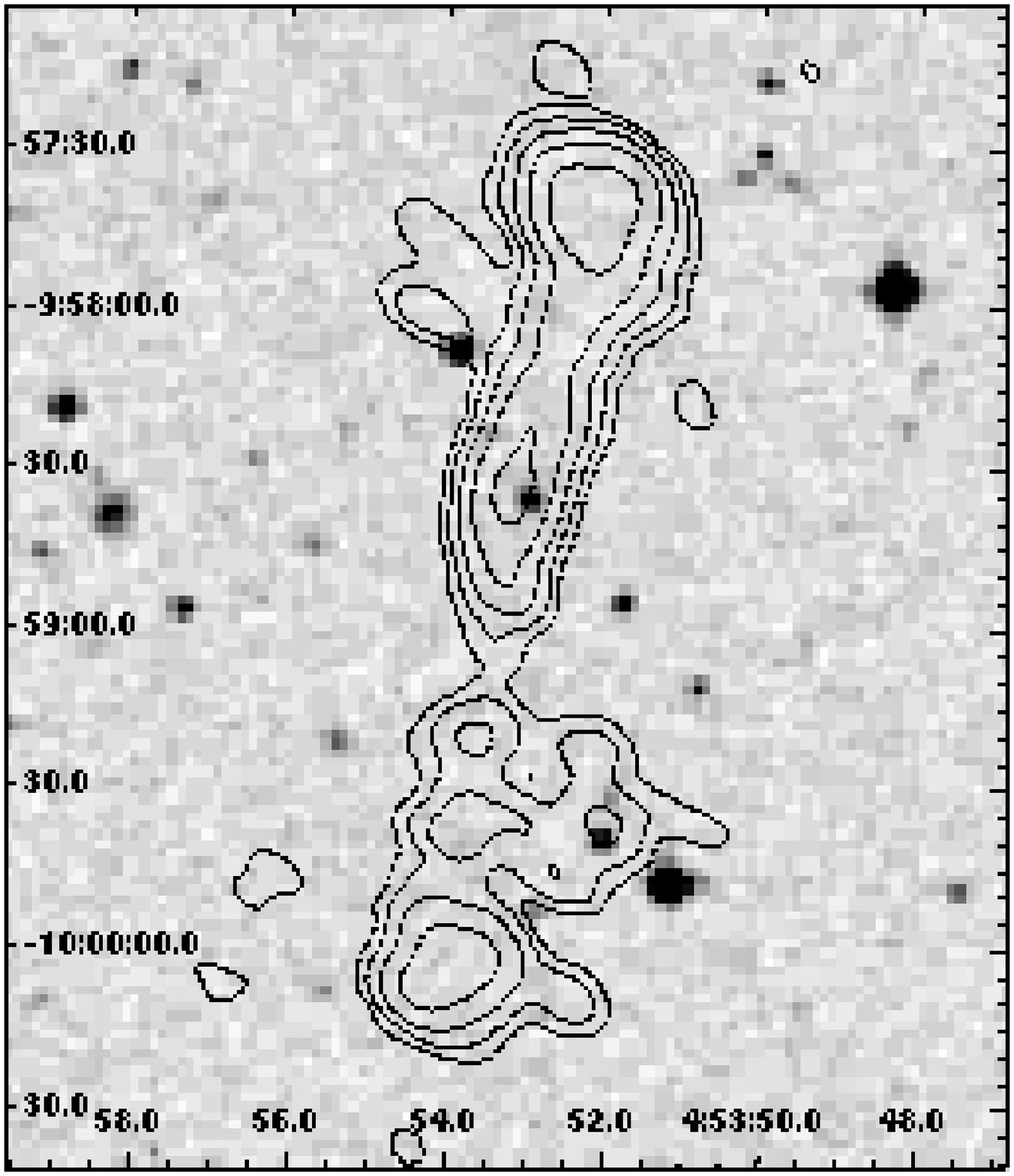}
\caption{610 MHz contours of the extended radio 
source J0453$-$0957, located at the North of \astrobj{A 521}, overlaid on the 
DSS--1 optical image. The radio contours
are 0.12 $\times$(-1, 1, 2, 4, 8, 16, 32, 64, 128, 
256, 512) mJy b$^{-1}$. The resolution is $13.1^{\prime \prime} 
\times 8.1^{\prime \prime}$, in p.a. $56^{\circ}$.}
\label{fig:sourcenorth}
\end{figure}

The radio galaxy population in \astrobj{A 521} is dominated by point--like
low power objects. However, two extended radio galaxies are well
visible in the northern region of the $30^{\prime}\times 30^{\prime}$ field 
of Figure \ref{fig:field}. Their radio emission is overlaid on the DDS--1
in Figs. \ref{fig:WAT} (J0454--1006) and \ref{fig:sourcenorth} (J0453--0957).
\\
Unfortunately, no redshift information is
available for the two optical counterparts, whose apparent magnitude is 
similar to that of the brightest cluster galaxies. Their (B$-$I)
colour, derived from the SuperCOSMOS catalogue,
is in very good agreement with the red sequence of the elliptical 
galaxies of the cluster, i.e. $<B-I> \sim 2.6$ (see Fig.15 in F03).
In particular, for 
J0454--1006 (B--I)=2.47 and for J0453--0957 (B--I)=2.64, therefore 
they might be part of \astrobj{A 521}, despite the large distance from the cluster
centre. 
If we assume that they are located at the average cluster distance, 
their total radio power is 
logP$_{610 \rm MHz}$= $25.26\pm0.08$ and logP$_{610 \rm MHz}$=$24.95\pm0.08$ 
for the wide--angle--tail J0454--1006 and the source J0453--0957 respectively.
These values are consistent with their radio structure, typical
of intermediate power radio galaxies.
\\
The tailed morphology of J0454--1006 is 
suggestive of interaction between the radio plasma and the external 
medium. Furthermore, even though this source is outside the 
boundary of the cluster X--ray emission (Figs. \ref{fig:location} and 
\ref{fig:relic1}), it is projected at a distance within the virial 
radius of \astrobj{A 521} (Tab. 1, Fig \ref{fig:field}). Therefore 
its distorted radio structure may be the result of a recent accretion 
at the cluster virial radius. On the contrary, the morphology of J0453--0957,
which is outside the virial radius, appears undisturbed.

\subsection{AGN activity in \astrobj{A 521}}\label{sec:RLF}

In order to understand if the ongoing merger event in \astrobj{A 521} has 
significant effect on the radio emission of the AGN cluster 
population, it is useful to compare the number of observed
radio galaxies in this merging environment with the number
expected from the the radio luminosity function (RLF) for 
early--type galaxies in normal clusters and in the field.
As reference we used the RLF at 1.4 GHz computed by 
Ledlow \& Owen 1996, hereinafter LO96) 
for early--type galaxies in a sample of local Abell clusters.
We are aware that the reference luminosity 
function was computed with a sample of clusters at lower redshift, 
i.e. z$\le$0.09. Unfortunately, no statistical information on the 
RLF of cluster radio galaxies at the redshift of \astrobj{A 521} is available
at present. This point will be further addressed in Section 7.
 
The analysis carried out by LO96 includes all radio galaxies 
within 0.3 Abell radius (R$_A$), with logP$_{1.4 \rm GHz}$ 
(W Hz$^{-1}$) $\ge$ 22.03 
and optical counterpart brighter than M$_R$=$-$20.5. 
This magnitude limit corresponds to m$_R$=19.8 at 
the distance of the cluster \astrobj{A 521}, and to the limit
I$_{\rm lim}$=19.0, adopting the (R$-$I) = 0.77 
colour for early--type galaxies at redshift z=0.2, 
reported in Fukugita et al. 1995).

For full consistency with LO96 we restricted our analysis to the
inner 0.3 R$_A$, which corresponds to 2.9 arcmin at the redshift
of \astrobj{A 521}.
The detection limit of our observations, i.e. 0.20 mJy,
corresponds to logP${1.4 \rm GHz}$ (W Hz$^{-1}$) = 22.14
in the cosmology of LO96 and assuming a spectral index 
$\alpha=0.8$ (Condon 1992). In order to avoid our incompleteness
in their first radio power bin, we compared the predictions from LO96 
and our detections for logP$_{1.4 \rm GHz}$ (W Hz$^{-1}$)$\ge$ 22.43,
which corresponds to a flux limit S$_{610 \rm MHz}$ $\simeq$ 0.36 mJy
(in our cosmology and again assuming $\alpha=0.8$).

Three early--type radio galaxies of \astrobj{A 521} match 
the above constraints for the radio power and the 
optical magnitude, i.e. $\sim 4.7\%$ of the total.

The total number of early--type galaxies belonging
to \astrobj{A 521} can be derived using the information in F03. 
A fraction of $57.6\%$ of their spectroscopic sample  
is composed by early--type galaxies, and for I $<$ I$_{lim}$, 
$85.5\%$ of these objects are cluster members
(see Figure 2 in F03). 
We corrected the number of early--type cluster galaxies for
the incompleteness of the F03 spectroscopic sample, which is
$\sim 55\%$ up to I$_{lim}$. We obtained that the total number 
of galaxies in the \astrobj{A 521} region brighter than I$_{lim}$ is 131, 
of which 112 belonging to the cluster. 
Taking all these constraints into account, we end up
with 64 early--type members in the inner 0.3 R$_A$ of \astrobj{A 521}.

On the basis of LO96, the expected number of radio emitting
early--type galaxies is 6, i.e. $\sim 10 \%$ of the total.

Allowing for the large poissonian uncertainties given by the small
numbers we are dealing with, our 3 detections are consistent
with the expectations from LO96 well within 1$\sigma$.

In Sect. 5.1 we pointed out the preferred location of the
radio galaxies along the NW--SE axis. Unfortunately, the incomplete 
optical information (i.e. non--uniform coverage of \astrobj{A 521}, see F03) 
does not allow any consideration on the connection between
the distribution of the radio galaxies and that of the 
early--type objects in the cluster.

\section{The relic source in \astrobj{A 521}}\label{sec:relic}

\begin{figure*}
\centering
\includegraphics[angle=0,width=14cm]{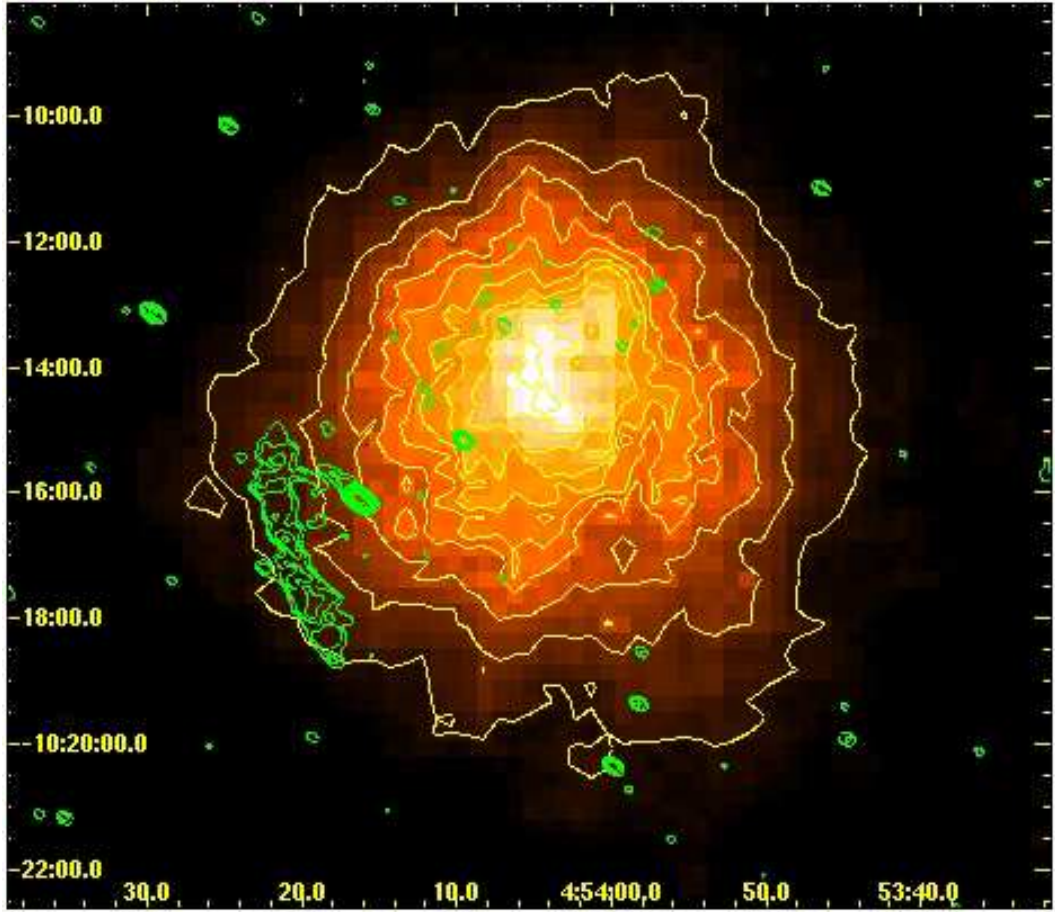}
\caption{610 MHz radio contours (green) of the \astrobj{A 521} region, 
overlaid on the X--ray smoothed ASCA image (colours and yellow contours) 
of the cluster, extracted from the ASCA public archive. The
resolution of the radio image is 
13.1$^{\prime \prime} \times$ 8.1$^{\prime \prime}$,
in p.a. $56^{\circ}$. The first radio contour is the 
5$\sigma$=0.2 mJy b$^{-1}$ level of the radio image.
The X--ray contours are 
3$\times$10$^{-5} \div 2.4 \times 10^{-4}$ cts s$^{-1}$ and
are spaced of 1.5 $\times 10^{-5}$ cts s$^{-1}$.}
\label{fig:relic1}
\end{figure*}

The most remarkable feature of \astrobj{A 521} 
is the presence of a region of diffuse radio emission
(Fig. \ref{fig:relic2}) in the south--eastern peripheral 
part of the cluster, at a projected distance of 
$\sim 4$ arcmin (i.e. 930 kpc) from the \astrobj{A 521} centre,
and at the border of the X--ray emission of the cluster 
(Figure \ref{fig:relic1}).

%%%%%%%%%%%%%%%%%%%%%%%%%%%%%%%%%%%%%%%
\begin{table} 
\caption[]{Properties of the relic source in \astrobj{A 521}.}
\smallskip
\begin{center}
\begin{tabular}{lc}
\hline\noalign{\smallskip}
S$_{610\rm MHz}$ (mJy) & 41.9$\pm$2.1 \\
S$_{1400 \rm MHz}$ (mJy) & 16.2$\pm$1.5 \\
$\alpha_{610}^{1400}$ &  1.14$\pm$0.16  \\
Linear size (kpc$\times$kpc) & $\sim$930 $\times$ 200 \\
axial ratio & $\sim$ 4.5 \\ 
d (arcmin)$^1$ & $\sim$ 4 \\ 
P$_{610\rm MHz}$ (10$^{24}$ W Hz$^{-1}$) &  8.13 \\
P$_{1400\rm MHz}$ (10$^{24}$ W Hz$^{-1}$) & 3.09\\
B$^{\prime}_{eq}$ ($\mu$G) $^2$&  1.3 \\
\hline
\end{tabular}
\end{center}
\label{tab:relic}
\hspace{1.5cm}$^1$ distance from the cluster centre

\hspace{1.5cm}$^2$ see Section 6
\end{table}
%%%%%%%%%%%%%%%%%%%%%%%%%%%%%%%%%%%%%%

The morphology of the source (labelled J0454--1017a in
Table 3) is arc--shaped and highly elongated.
Its total angular size along the major axis is
$\sim 4^{\prime}$, corresponding to a linear size of $\sim$ 930 kpc, 
and its largest transversal angular size is only $\sim 50^{\prime \prime}$,
corresponding to 200 kpc.
This source was first detected at 1.4 GHz with the Very Large Array
(VLA) by Ferrari (2003), and an image is given also in the appendix 
of F05.

The resolution of the radio image in Figs. \ref{fig:relic2} 
and \ref{fig:relic1} is high enough to rule out the possibility
that this object is a blend of different radio sources. 
If we exclude the cluster radio galaxies 
embedded in the diffuse emission (the point sources A, B and C 
in Fig. \ref{fig:relic2} and Table 4), 
the extended radio source does not appear to be associated 
with any optical counterpart. 

The size and the radio morphology, as well as the 
lack of an optical identification, suggest that 
the diffuse radio source located at the outskirts of 
\astrobj{A 521} can be included in the class of cluster relics. 
\\
Fig. \ref{fig:relic2} 
also shows that the cluster radio galaxy A (J0454--1016a, the most powerful 
source in \astrobj{A 521}) is located only $1.5^{\prime}$ away
from the relic (in the plane of the sky), and a faint bridge of radio
emission connects the two sources. Even though projection effects in
\astrobj{A 521} should be taken into account, we note that this situation is similar to 
what is found in the Coma cluster, where a bridge of radio emission connects the 
tails of the radio galaxy \astrobj{NGC 4789} and the prototype relic source 
\astrobj{1253+275} (Giovannini et al. 1991).

In order to properly determine the value of the 
total flux density of the relic at 610 MHz, we  
integrated over the whole region covered by its emission
and subtracted the flux density of the embedded
point--sources (see Section 4). The flux density is 
S$_{610MHz}$= 41.9$\pm$2.1 mJy, which gives a radio 
power logP$_{610MHz}$ (W Hz$^{-1}$) = 24.91.

Using the NVSS information, we estimated the total spectral 
index of the relic between 610 MHz and 1400 MHz.
The NVSS flux density at 1400 MHz is 
S$_{1400MHz}$=16.2$\pm$1.5 mJy (after subtraction of the 
embedded point sources), and therefore
$\alpha_{610MHz}^{1400MHz} = 1.14 \pm 0.16 $.

A zero--order estimate of the energy density of the relativistic plasma and
magnetic field associated with the relic can be obtained under the assumption of
minimum energy conditions (e.g. Pacholczyk 1970). 
Assuming a power law spectrum for the electrons  with slope
$\delta=2\alpha + 1$ ($\alpha=1.14$), and the classical minimum energy
equations (normally computed in the frequency range between 
$\nu_1$=10 MHz and $\nu_2$=100 GHz), we obtained an equipartition 
magnetic field B$_{\rm eq}$=0.4 $\mu$G.
However, we note that for this value of $B_{\rm eq}$, the electrons with  
Lorentz factor $\gamma \sim 2.5\times10^3$ emit at 10 MHz, and thus the contribution
of electrons with $\gamma < 2.5 \times 10^3$ to the total energy density
is not taken into account.
A more accurate approach is given by adopting equipartition equations 
with a low energy cut--off $\gamma_{\rm min}$ in the electron energy
distribution (not in the emitted synchrotron spectrum). Using the
equations given in Brunetti et al. 1997), we derived the
following value for the magnetic field:

$${\rm B_{eq}^{\prime} \simeq 1.3 \times (\frac{\gamma_{min}} {50})^
{\frac{{1-2\alpha}} {\alpha +3}}}$$

\noindent with ${\rm B_{eq}}^{\prime}$ expressed in $\mu$G.
The parameters of the relic source are given in Table \ref{tab:relic}
(${\rm B_{eq}}^{\prime}$ given for $\gamma_{min} = 50$).

Both values of the equipartition magnetic field are in agreement
with the estimates found in the literature for radio relic sources
and cluster radio halos, i.e. in the range 0.1 -- 1 $\mu$G 
(see the review by Govoni \& Feretti 2004).

\section{Discussion}\label{sec:discussion} 

The main results of our 610 MHz GMRT study of
\astrobj{A 521} can be summarized as follows.

\parn i) We detected a relic source, whose projected location
is just at the boundary  of the X--ray emission from the 
intracluster gas (Section 6);

\parn ii) We compared the number of detected radio loud AGN with 
the expectations from the radio luminosity function by LO96
(inner 0.3 R$_A$ and logP$_{\rm 1.4~GHz}$(W Hz$^{-1}) \ge$ 22.43)
and found 3 objects, to be compared to the 6 expected (Section 5.3).
\\
\\
Point (i) is by far the most relevant.
In the following we will discuss these results 
in the light of the assessed ongoing merger in this galaxy cluster.

\subsection{Cluster merger and AGN/starburst radio activity}  

Our analysis on the cluster radio galaxies showed that
the number of radio emitting early--type galaxies in
the \astrobj{A 521} is consistent
with the expectations from the standard RLF
(3 against 6) if we allow for the large uncertainties 
due to the very small number statistics.
Such comparison should be taken with care, since
it is done with the local (z$\le$0.09) radio luminosity function 
for cluster ellipticals (see Section 5.3), 
while \astrobj{A 521} is at redshift 0.247. 
However, this result is striking if we take into account the positive 
evolution of the RLF for X--ray selected high redshift clusters 
(Stocke et al. 1999, Branchesi et al. 2005), and leads us to
safely conclude that the multiple merger events in \astrobj{A 521} are not
increasing the probability of an early--type galaxy to develop
a nuclear radio source compared to other less extreme environments,
as already found in the complex merging
environment of \astrobj{A 3558}, in the central region of the Shapley
Concentration (Venturi et al. 2000; Giacintucci et
al. 2004).

The optical analysis of F03 showed that the late--type star forming
galaxies, are preferentially located on the axis perpendicular to the 
direction of the merger (see Fig. 1).
The radio power limit in our observations 
(logP$_{610 \rm MHz}$ (W Hz$^{-1}$)= 22.54)
favours the detection of AGNs, so very little can be inferred on the
role of the ongoing merger on the starburst activity in this cluster.
However, if we consider logP$_{610MH}$ (W Hz$^{-1}$) = 23 as
a reasonable upper limit for the radio emission 
from starburst galaxies (e.g. Condon 1992), 
we can conclude that no major starburst emission is detected. 
Unfortunately, no information on the infrared flux of the late--type
radio galaxies in our sample is available in the literature to give support 
to our conclusions.

\subsection{The merging events in \astrobj{A 521}  and the formation of the relic}  

The most important result of this paper is the detection of the
relic source, which gives further observational support to the
hypothesis of a close connection between cluster mergers and
relic radio emission. 
The relic in \astrobj{A 521} is located in projection at the border of the 
cluster X--ray emission. It is slightly inclined with 
respect to the outer ring of the ASCA X--ray isophotes (see
Fig. 8). 

A number of models have been proposed to explain 
the origin of radio relics. All these models invoke a 
connection between these sources and the presence of a 
shock within the X--ray gas driven by a merging episode
(see for instance Markevitch et al. 2005). 
Simulations of cluster mergers (Ricker \& Sarazin, 2001)
show that the merging of two subclusters leads to the formation 
of two shocks (front and back shock). 

\astrobj{A 521} has a complex dynamics. The main cluster has a mass
of the order of 10$^{15}$ M$_{\odot}$ and has been undergoing
multiple minor merging events with groups whose mass is $\sim$ 1/10
lower. Using the dynamical analysis of M00, F03 and F05, and from
inspection of Figure 1, a possible scenario is that 
the group G111 is falling onto the main cluster G11,
coming from North--West. Furthermore, the presence of the two optical 
groups G12 and G112 south of the central part of the cluster 
(see Figs. 1 and 3) suggests that also the southern part of the cluster 
region may be dynamically active.

\subsubsection{Merger shock}

One possibility is that relativistic electrons are accelerated 
from the thermal pool by the passage of a strong merger shock 
(Ensslin et al. 1998, R\"ottgering et al. 1997). In this case the spectrum
of the emitting electrons\footnote{We do not consider the case
of reacceleration of a pre-existing population of relativistic
electrons (see Markevitch et al. 2005 for a detailed discussion).} 
is related to the Mach number $M$ of the
shock by 

$$\delta=2\frac{(M^2+1)}{(M^2 -1)} + 1$$

\noindent (e.g. Blandford \& Eichler 1987). Here we include also
the effect of particle aging $\Delta(\delta)=1$, which comes out
from the combined effect of Inverse Compton energy losses and
continuous injection.
In the case of \astrobj{A 521}  the spectral index of the relic is $\alpha=1.14$, 
which gives $\delta=3.28$, and the requested Mach number of the shock
is $M\sim 3.9$.

\subsubsection{Adiabatic compression}

A second possibility is adiabatic compression of fossil radio plasma by 
a merger shock (Ensslin \& Gopal--Krishna 2001, hereinafter EG--K01).
In this case, the numerical 3--D MHD simulations by Ensslin \& Br\"uggen
(2002, hereinafter EB02) predict a variety of radio
morphologies and polarization properties, which may be reasonably 
well matched by the available high sensitivity radio images.
Another important requirement in this scenario 
is the presence of an active radio galaxy in the proximity of the
relic. This constraint is satisfied in \astrobj{A 521}, where the radio galaxy 
J0454--1016a is located only 1.5$^{\prime}$ from the relic, i.e. 
a projected distance of $\sim$ 350 kpc.
A previous cycle of activity of this radio galaxy could have provided 
the fossil radio plasma in the ICM, revived by the shock compression.

\subsubsection{Ram pressure stripping}

Finally, another appealing possibility is that the relic in \astrobj{A 521} is the 
result of ram pressure stripping of the radio lobes of J0454$-$1016a 
{\it (a)} by group merger in the southern cluster region, or 
{\it (b)} by the infalling of J0454$-$1016a itself through G11. 
\\
This scenario requires that the internal pressure of the lobes $P_{int}$
is significantly smaller than the external ram pressure, i.e.

$$P_{int} \approx 2 \frac{B_{eq}^{'2}}{8 \pi} << \rho_{\rm ICM} v^2_{\rm merg} \approx 10^{-12} 
n ~ v^2_{\rm merg}$$

\noindent where $v_{\rm merg}$ is the infalling velocity in units 
of 1000 km s$^{-1}$,
$\rho_{\rm ICM}$ is the density of the intracluster medium,  
and $n$ is the number density in units of $10^{-4}$ particle cm$^{-3}$. 
\\
The projected distance of J0454$-$1016a from the relic 
($\sim 350$ kpc) requires a time of the order of 

$$t_{cross} \approx 3.5 \times 10^8 
(\frac{v_{merg}}{10^3 {\rm km/s}})^{-1}~~~~ {\rm yr}$$

\noindent 
to be crossed by the ICM of any merging group
(case {\it (a)}), or  by J0454--1016a itself (case {\it (b)}).  
We note that in {\it (a)} we assume that the core position of 
J0454--1016a is not affected by the group merger dynamics.
In order to allow the electrons in the radio lobes to still
emit in the radio band, the time $t_{cross}$ should be smaller 
than the life--time of the radiating electrons. This implies 
$v_{merg}$ \gtsim $~$3000 km/s. Such velocity leads to 
a Mach number $M$ \gtsim$~$ 2 for the merging group, 
or for J0454--1016a.

\subsection{Is there a shock in the external region of \astrobj{A 521}?}

All the possibilities given above to explain the formation of
a radio relic require the presence of a shock in the external
region of \astrobj{A 521}, with Mach number in the range 
2 \ltsim $M$ \ltsim $~$4. 
In search for an observational signature of a shock in the relic region
we re--analysed the public archive Chandra ACIS--I (39 ksec, OBSID 901)
and ACIS--S (39 ksec OBSID 430) observations, analysed also in F05.
We point out that the work in F05 is mainly restricted to the central
and northern part of the cluster, and does not include the region which is
relevant to the present discussion.
 
We processed tha data using CIAO 3.2 and the newest calibration
database CALDB 3.1.0. 
On the (0.5--5) keV background--subtracted 
and exposure--corrected image of the cluster, we extracted the 
radial X--ray surface brightness profile perpendicular to the relic source,
using a 80$^{\circ}$ sector centered on the cluster centre and containing
the relic. 
\\
The X--ray brightness profile does not show any evidence in 
support of the existence of a shock front at the projected location of 
the relic. This result is consistent with what we found in 
re--analysing the archive ASCA data (see also Fig. 8).
\\
We point out, however, that this is not enough to rule out a connection 
between the relic and the presence of a shock. Two more issues
should be considered. In particular:
{\it (a)} the relic is very peripheral, therefore the cluster X--ray surface 
brightness is very low here, and deeper X--ray imaging is necessary to 
investigate the presence of a shock; {\it (b)} projection effects 
should also be taken into account in the analysis. We defer a detailed 
discussion on this point to a future paper.

\section{Summary}\label{sec:summary}

In this paper we presented high sensitivity observations of the
cluster of galaxies \astrobj{A 521}, carried out at 610 MHz with the GMRT.
The cluster is known to have a complex dyamics, and 
the radio emission from the cluster was analysed in detail, 
using the multiband (X--ray and optical) information available in 
the literature.

We found that the AGN activity in the cluster is consistent
with the local RLF for cluster ellipticals within the 
large poissonian errors, i.e. we have 3 detections out of 6 expected
sources. This result suggests that 
the multiple merging events in \astrobj{A 521} are not increasing the AGN
radio activity in the early--type population compared to other
environments.

A radio relic was detected at the cluster periphery, and a 
few possible scenarios for the presence of this relic were 
discussed.
\\
One possibility is that the relic is connected to the 
presence of shock waves induced by the merger.
Such shocks may have accelerated relativistic electrons or ``revived'' 
fossil radio plasma 
through adiabatic compression of the magnetic field or shock re--acceleration. 
The presence of an active cluster radio galaxy in the proximity of
the relic suggests that the ``revived'' plasma might be connected to 
previous cycles of activity in this object. The radio properties of
the relic require a high Mach number for such shock. 
\\
Another possible explanation is ram pressure stripping of radio
lobes associated with the nearby radio galaxy J0454--1016a as a
consequence of group infalling/merger. We showed that 
the projected distance between the relic and the radio galaxy 
may be in reasonable agreement with the age of the radiating electrons 
in such sources.
\\
An analysis of public archive Chandra data does not provide
observational evidence for the presence of a shock at the location of
the relic, however projection effects should be taken into account
for an accurate study.
\\
\\
We thank the staff of the GMRT for their help during the observations.
GMRT is run by the National Centre for
Radio Astrophysics of the Tata Institute of Fundamental Research.
G.B. and  R.C. acknowledge partial support from MIUR from grant 
PRIN2004. The authors wish to thank P. Mazzotta for insightful discussions
and help with the analysis of the X--ray archive data.

% \appendix

% \section{}
% \label{}

% Bibliographic references with the natbib package:
% Parenthetical: \cite{Bai92} produces (Bailyn 1992).
% Textual: \citet{Bai95} produces Bailyn et al. (1995).
% An affix and part of a reference:
%   \cite[e.g.][Ch. 2]{Bar76}
%   produces (e.g. Barnes et al. 1976, Ch. 2).

%\begin{thebibliography}{}

%\bibitem[Names(Year)]{label} or \bibitem[Names(Year)Long names]{label}.
% (\harvarditem{Name}{Year}{label} is also supported.)
\vskip 1 truecm

{\bf References}

\par\medskip\noindent
Arnaud M., Maurogordato S., Slezak E., Rho J.,
2000, A\&A, 355, 461 (A00)
\par\medskip\noindent
Bagchi J., Durret F., Neto G., Surajit P., Chavan S., 
2005, 29th International Cosmic ray Conference, Pune, 
in press (astro--ph/0508013)
\par\medskip\noindent
Blandford R. \& Eichler D., 1987, Phys. Rep., 154, 1
\par\medskip\noindent
B\"ohringer H., Schuecker P., Guzzo L., Collins C. A.,
Voges W., Cruddace R. G., Ortiz-Gil A., Chincarini G.,
De Grandi S., Edge, A. C, 2004, A\&A, 425, 367
\par\medskip\noindent
Branchesi M., Gioia I.M., Fanti C., Fanti R., Perley, R.A., 
2005, A\&A, in press (astro--ph/0509138)
\par\medskip\noindent
Brunetti G., Setti G., Comastri A., 1997
A\&A, 325, 898
\par\medskip\noindent
Brunetti G.,
2003, in {\it Matter and Energy in Clusters of Galaxies}, 
ASP Conf. Series 301, p.349
\par\medskip\noindent
Brunetti G., in {\it Outskirts of galaxy clusters: intense life in the
suburbs}, 2004, IAU colloquium 195, Ed. A. Diaferio,Cambridge
Univ. Press,  p.148
\par\medskip\noindent
Buote D.A.,
2001, ApJ, 553, L15
\par\medskip\noindent
Cassano R., Brunetti G., Setti G., 2004, JKAS, 37, 589
\par\medskip\noindent
Cassano R., Brunetti G., 2005, MNRAS, 357, 1313
\par\medskip\noindent
Condon J.J., 1992, A\&A Rev., 30, 575
\par\medskip\noindent
Condon J.J., Cotton W.D., Greisen E.W, et al., 1996, ADIL, JC, 01
\par\medskip\noindent
Ensslin T.A., Biermann P.L., Klein U., Kohle S.,
1998, A\&A 332, 395
\par\medskip\noindent
Ensslin T.A., Br\"uggen M., 2002, MNRAS, 331, 1011 (EB02)
\par\medskip\noindent
Ensslin T.A., Gopal--Krishna, 2001, A\&A, 366, 26 (EG--K01)
 \par\medskip\noindent
Feretti L., Orr\'u E., Brunetti G., Giovannini G., Kassim N.,
Setti G., 2004, A\&A, 423, 111
\par\medskip\noindent
Feretti L., Fusco--Femiano R., Giovannini G., Govoni G.,
2001, A\&A, 373, 106
\par\medskip\noindent
Feretti L., 2005, in {\it X-ray and Radio Connections},
eds. L.O. Sjouwerman \& K.K Dyer,
astro-ph/0406090
\par\medskip\noindent
Ferrari C., 2003, Ph.D. Thesis, University of Nice--Sophia Antipolis,
France
\par\medskip\noindent
Ferrari C., Arnaud M., ettori S., Maurogordato S.,
Rho J., 2005, A\&A, in press (astro--ph/0508585)
(F05)
\par\medskip\noindent
Ferrari C., Maurogordato S., Cappi A., Benoist C.,
2003, A\&A, 399, 813 (F03)
\par\medskip\noindent
Fukugita, M., Shimasaku, K., \& Ichikawa, T.,
1995, PASP, 107, 945
\par\medskip\noindent
Giacintucci S., Venturi T., Bardelli S., Dallacasa D., Zucca E., 
2004, A\&A, 419, 71 
\par\medskip\noindent
Giacintucci S., Venturi T., Brunetti G., Bardelli S., Dallacasa D., Ettori S.,
Finoguenov A., Rao A.P., Zucca E., 
2005, A\&A, 440, 867
\par\medskip\noindent
Giovannini G., Feretti L., Stanghellini C., 1991, A\&A, 252, 528
\par\medskip\noindent
Giovannini G., Feretti L., 2002, in 
{\it Merging Processes in Galaxy Clusters}, 
ed. L. Feretti, I.M. Gioia, G. Giovannini,
ASSL, 272, 197
\par\medskip\noindent
Giovannini G., Feretti L.,
2004, JKAS, 37, 323
\par\medskip\noindent
Govoni F., Feretti L., 
2004, Journal of Mod. Phys., Vol. 13, Issue 8, p. 1549
\par\medskip\noindent
Hambly, N. C., MacGillivray, H. T., Read, M. A., Tritton, S. B.,
Thomson, E. B., Kelly, B. D., Morgan, D. H., Smith, R. E.,
Driver, S. P., Williamson, J., Parker, Q. A., Hawkins, M. R. S.,
Williams, P. M., Lawrence, A.
2001, MNRAS, 326, 1279
\par\medskip\noindent
Kempner J.C., Sarazin C.L., 2001, ApJ, 548, 639
\par\medskip\noindent
Ledlow M.J. \& Owen F.N.,
1996 ApJ, 112, 9 (LO96)
\par\medskip\noindent
Maddox S.J., Efstathiou G., Sutherland W.J., Loveday J.
1990, MNRAS, 243, 692
\par\medskip\noindent
Markevitch M., Govoni F., Brunetti G., Jerius D.,
2005, ApJ, 627, 733
\par\medskip\noindent
Maurogordato S., Proust D., Beers T.C., Arnaud M., Pell\'o R., Cappi A.,
Slezak E., Kriessler J.R.,
2000, A\&A, 355, 848, (M00)
\par\medskip\noindent
Miller N.A., Owen F.N., Hill J.M., 2003,
AJ, 125, 2393
\par\medskip\noindent
Miller N.A., Owen F.N., 2003, AJ, 125, 2427
\par\medskip\noindent
Owen F.M., Ledlow M.J., Keel W.C., Morrison G.E.,
1999, AJ, 118, 633
\par\medskip\noindent
Pacholczyk A.G., {\it Radio Astrophyisics}, 1970, 
Freeman Eds.
\par\medskip\noindent
Prandoni I., Gregorini L., Parma P. et al. 2000,
A\&AS, 146, 41
\par\medskip\noindent
Reiprich T.H. \& B\"ohringer H., 2002,
ApJ, 567, 716
\par\medskip\noindent
Ricker P. M. \& Sarazin C. L.,
2001, ApJ, 561, 621
\par\medskip\noindent
Roettiger K., Burns J.O., Stone J.M., 1999,
ApJ, 518, 603
\par\medskip\noindent
R\"ottgering H.J.A, Wieringa M.H., Hunstead R.W., 
Ekers R.D., 1997, MNRAS 290, 577
\par\medskip\noindent
Sarazin C.L., 2002, in
{\it Merging Processes in Galaxy Clusters}, 
ed. L. Feretti, I.M. Gioia, G. Giovannini,
ASSL, 272, 1
\par\medskip\noindent
Sarazin C.L., 2004, 
in {\it X--ray and radio connections meeting},
ed. L. O. Sjouwerman \& K. K. Dyer, 
astro--ph/0406181
\par\medskip\noindent
Schlegel D.J., Finkbeiner D.P., Davis M., 1998,
ApJ, 500, 525
\par\medskip\noindent
Schuecker P., Bhringer H., Reiprich T.H., Feretti L.,
2001, A\&A, 378, 408
\par\medskip\noindent
Stocke J.T., Perlman E.S., Gioia I.M., Harvanek M., 1999,
AJ 117, 1967
\par\medskip\noindent
Unewisse A.M., Hunstead D.W. \& Pietrzynski B., 
1993, Publ. Astron. Soc. Austr., 10, 229
\par\medskip\noindent
Venturi T., Bardelli S., Morganti R., Hunstead R.W.,
2000, MNRAS, 314, 594 
\par\medskip\noindent
Venturi T., Bardelli S., Dallacasa D., Brunetti G.,
Giacintucci S., Hunstead R.W., Morganti R.,
2003, A\&A, 402, 913

%\end{thebibliography}

\end{document}